\documentclass[twocolumn,superscriptaddress,floatfix,longbibliography,aps,footinbib]{revtex4-1}
\usepackage{graphicx}
\usepackage{amsmath}
\usepackage{amssymb}
\usepackage{xspace}

\newtheorem{thm}{Theorem}[section]
\newtheorem{thmmain}{Theorem}
\graphicspath{{figures/}}

\usepackage[]{hyperref}
\hypersetup{
  colorlinks   = true,
  urlcolor     = blue,
  linkcolor    = blue,
  citecolor   = red
}

\newcommand{\myhead}[1]{\textit{{#1}.---}}

\newcommand{\figref}[1]{Fig.~\ref{#1}}
\newcommand{\secref}[1]{Sec.~\ref{#1}}

\renewcommand{\d}[1]{\ensuremath{\operatorname{d}\!{#1}}}

\newcommand{\ilatrv}[2]{{\mathcal{L}}^{-1}_{{#1}}\left\{{#2}\right\}}

\newcommand{\latr}[1]{{\mathcal{L}}\left\{{#1}\right\}}

\newcommand{\latrv}[2]{{\mathcal{L}}_{{#1}}\left\{{#2}\right\}}

\newcommand{\lt}{\hat}

\newcommand{\ft}{\tilde}

\newcommand{\flt}[1]{{\hat{\tilde #1}}}

\newcommand{\partialt}[1]{{\partial_{t}{#1}}}

\newcommand{\partialxx}[1]{{\partial_{xx}{#1}}}

\newcommand{\Dgen}[1]{\mathcal D_{#1}}

\newcommand{\Da}{\Dgen{\alpha}}

\newcommand{\ts}{\tau_\text{m}}

\newcommand{\lm}{\delta_\text{m}}

\newcommand{\Dt}{\Delta}

\newcommand{\lv}{\lambda}

\newcommand{\resdens}{p_\text{r}}

\newcommand{\ensav}[1]{\left\langle #1 \right\rangle}

\newcommand{\pDt}{p_{\Dt}}

\newcommand{\cst}{c_{\text{s}}}

\newcommand{\prst}{p_{\text{r,s}}}

\newcommand{\Xt}{X}

\newcommand{\Wp}{Y}

\newcommand{\Xtr}{X_{\text{r}}}

\newcommand{\xc}{{\mathbf{x}}}

\newcommand{\TD}{R}

\newcommand{\pTD}{p_{\TD}}

\newcommand{\TDc}{R_{\text{d}}}

\newcommand{\pTDc}{p_{\TDc}}

\newcommand{\kRc}{k_{\TDc}}

\newcommand{\kR}{k_{\TD}}

\newcommand{\Kda}{{K}}

\newcommand{\rt}{T}

\newcommand{\rn}{N}

\newcommand{\Ind}[1]{\mathbb I({#1})}

\newcommand{\Fr}{F _{\text{r}}}

\newcommand{\Fp}{F_{\text{p}}}

\newcommand{\Fst}{F_{\text{r,s}}}

\newcommand{\Fs}{F_{\text{s}}}

\newcommand{\fr}{f_\text{r}}

\newcommand{\tshift}{\tau}

\newcommand{\fq}{\omega}

\newcommand{\gf}{g}

\newcommand{\rex}{\nu}
\newcommand{\bex}{\beta}

\newcommand{\CFPS}{diffusion under disordered reset\xspace}
\newcommand{\CCFPS}{Diffusion under disordered reset\xspace}

\newcommand{\csicaf}{\affiliation{Spanish National Research Council (IDAEA-CSIC), E-08034 Barcelona, Spain}}

\begin{document}

\newcommand{\figw}{ {1.0\linewidth} }

\title{Unified approach to reset processes and application to coupling between process and reset}

\author{G. J. \surname{Lapeyre}, Jr.}
\csicaf
\author{M. Dentz}
\csicaf

\date{\today}

\begin{abstract}
  We present a unified approach to those observables of stochastic processes under reset
  that take the form of averages of functionals depending on the most recent renewal
  period.
  We derive solutions for the observables, and determine the conditions
  for existence and equality of their stationary values with and without reset.
  For intermittent reset times, we derive exact asymptotic expressions
  for observables that vary asymptotically as a power of time.
  We illustrate the general approach with general and particular results for the
  power spectral density, and moments of subdiffusive processes.
  We focus on coupling of the process and reset
  via a diffusion-decay process with microscopic dependence between transport and decay.
  In contrast to the uncoupled case,
  we find that restarting the particle upon decay does not produce a probability current
  equal to the decay rate, but instead drastically alters the time dependence
  of the decay rate and the resulting current.
\end{abstract}

\maketitle 

The action of interrupting a process and restarting
it afresh from its initial state has found wide
application in mathematics, physical sciences,
and beyond~\cite{Levikson1977,Pakes1978,Pakes1981,Pakes1997,Kyriakidis1994,Economou2003,Evans2011,%
  Evans2011a,Evans2013,Montero2013,Evans2014,Christou2015,Pal2015,Majumdar2015,Pal2016,%
  Nagar2016,Eule2016,Montero2017,Chatterjee2018,Majumdar2018,Hollander2018,Manrubia1999,Janson2012}.
Reset processes appeared in
mathematical studies of Markov chains that ``return'',
or undergo ``killing and resurrection''
by moving immediately
under some condition
to the initial
state~\cite{Levikson1977,Pakes1978,Pakes1981,Pakes1997,Kyriakidis1994}.
More recently, diffusion with random resetting (or restarting) has played a seminal and ongoing role
in furthering our understanding of
processes under reset~\cite{Evans2011,Evans2011a,Evans2013,Montero2013,Evans2014,Gupta2014,Christou2015,Pal2015,Majumdar2015,%
  Pal2016,Nagar2016,Eule2016,Montero2017,Harris2017,Boyer2017,Falcao2017,Roldan2017,Chatterjee2018,Majumdar2018}.
A prime application is first passage~\cite{Redner2001,Metzler2014a} under reset
applied to search strategies~\cite{Benichou2007,Evans2013,Kusmierz2014,Eliazar2017,Chechkin2016,Chupeau2017,Evans2018,Pal2019},
such as animal foraging~\cite{Benichou2011}, biochemical reagents locating
targets~\cite{Eliazar2007}, and completion of
reactions~\cite{Durang2014,Roldan2016,Reuveni2014,Rotbart2015}.

A major focal point in these works
is the effect of restarting on occupation distributions,
which typically approach a non-equilibrium stationary distribution.
An early prototypical example is
Brownian motion reset at constant rate~\cite{Evans2011}.
Since then, a wide variety of directions have been explored,
including fluctuating interfaces~\cite{Gupta2014},
Levy flights~\cite{Kusmierz2014},
non-Poissonian reset times~\cite{Eule2016,Nagar2016},
time-dependent resetting~\cite{Pal2016}
and diffusion~\cite{BodrovaA2018,Bodrova2018},
residence times~\cite{MasoPuigdellosasA2019,MasoPuigdellosas2019},
bounded domains~\cite{Christou2015,Chatterjee2018},
potential landscapes~\cite{Pal2015},
interacting particles~\cite{Falcao2017},
quantum dynamics~\cite{Rose2018,Mukherjee2018},
and continuous time random walks~\cite{Montero2013,Montero2017,Shkilev2017,Kusmierz2018,dosSantos2019}.
Recent research has moved beyond distributions to other other observables
including spectral density~\cite{Majumdar2018}, and additive functionals~\cite{Meylahn2015,Hollander2018,PalA2019}.

All of these studies have assumed an extrinsic reset mechanism,
independent reset and transport, and either a homogeneous reset,
or one with deterministic spatial variation~\cite{Evans2011a}.
In particular, they neglect the effects of disorder.
In fact, disorder is ubiquitous, and processes on disordered media
import in engineering~\cite{Sahimi1994},
sciences~\cite{Haus1987,Bouchaud1990,Stauffer1994,Havlin2002,Barkai2012,Edery2014,Hunt2014},
and mathematics~\cite{Grimmett1990, Grimmett1999b}.
In the context of searching, imagine a scenario in which local random variations
influence the decision whether to stop searching and
restart as well as the rate at which the environment is explored.
Mathematically, in the presence of disorder,
fluctuations in the local environment
couple the transitions and reset and have a profound
effect on the reset rate.

The expanding diversity of
applications of reset processes
has been accompanied by growing acknowledgment of the need for
unifying approaches to disparate results.
Fundamental derivations are repeated (See references in Ref.~\cite{Meylahn2015}),
using a variety of techniques and including details of particular processes
which obscure the essence of basic principles.
Efforts to fill this need go back to restarted discrete Markov processes~\cite{Pakes1981,Pakes1997},
and continue in particle densities~\cite{Eule2016}
and first passage statistics~\cite{Reuveni2016, Pal2017, Chechkin2016}.
However, we still lack a unified framework for analyzing functionals
of a reset processes.
Ideally, such a framework would encompass not only models studied to date,
but also processes on disordered media, and, more generally, dependent processes and reset.

In this article we present a framework for
analyzing functionals of reset processes and apply it to
diffusion under decay on a disordered medium in which
small-scale fluctuations introduce dependence between
the diffusion and reset time.
In contrast to previous studies, we find that resetting the particle upon decay
does not simply counter-balance the decay rate, but rather drastically
alters the emergent decay/reset rate.
However, our framework goes much further in
that it may be easily applied to any stochastic process subject to any sequence
of i.i.d. random reset intervals.
In particular, we provide a number of fundamental and generic results
for processes under intermittent resetting.

\myhead{Functionals of stochastic processes under reset}%
%
Let $\Xt(t)$ be a stochastic process with no restrictions
on state space or transitions,
although for convenience,
we often refer to $\Xt(t)$ as the state
of a particle at time $t$.
The process under reset is obtained by repeatedly
restarting $\Xt(t)$ at random time intervals $\{\TD_{1},\TD_{2},\ldots\}$,
which are i.i.d. copies of the reset time $\TD$.
$\Xt$ and $\TD$ may be dependent.
The time of the $n$th reset $\rt_{n}$ is then determined by
$\rt_{1}=\TD_{1}$ and $\rt_{n+1} = \rt_{n}+\TD_{n+1}$.
More precisely, during the $n$th reset period,
the reset process is given by $\Xt_{n}(t-\rt_{n})$,
where $\{\Xt_{1},\Xt_{2},\ldots\}$, are i.i.d. copies of $\Xt(t)$.
The number of resets performed up to time $t$ is the renewal process
$\rn_{t}\equiv \max\{n: \rt_{n} \le t\}$.
%
%
At any time $t$, the
time of the most recent reset is $T_{\rn_{t}}$.
The restart process is then given
by $\Xtr(t) \equiv \Xt_{\rn_{t}}(t-T_{\rn_{t}})$.
We study the average of functionals
(meaning any function of a function) of the reset process
$\Fr(t)\equiv \ensav{f(\Xtr(t'), t)}$
that depend only on the most recent renewal period,
that is
on $\Xtr(t')$ for $T_{\rn_{t}} \le t' < T_{\rn_{t}+1}$.
Particular instances of $\Fr(t)$ may depend on additional parameters.
The crucial point is that the restriction to
dependence on the most recent reset period means that $\Xt(t)$
enters only via the quantity
$\Fp(t)$ defined by
\begin{equation}
  \Fp(t) = \ensav{f(\Xt, t)\Ind{t < \TD}}
  = F(t) \pTD(t),
  \label{Fpdef}
\end{equation}
where $F(t)\equiv \ensav{f(\Xt, t) \mid t < \TD}$,
and $f(\Xt,t)$
[where $\Xt$ implies $\Xt(t')$]
is the functional of the process without reset,
and $\pTD(t)\equiv\Pr(t<\TD)$ is the survival probability.
We emphasize that~\eqref{Fpdef} by itself does not describe a process under
reset, but rather under decay
(or removal).
The reset process is then obtained by restarting
the removed particle.
This class of functionals covers the vast majority of
quantities under reset studied to date and many more.
All one-point functions,
such as the density and its moments are included.
But, we also treat functionals of the entire renewal period.

A renewal calculation yields the equation relating
the averages under death $\Fp(t)$ and reset $\Fr(t)$
\begin{equation}
   \begin{aligned}
     \Fr(t)
     = \Fp(t)
     +  \int_{0}^{t} \kR(t') \ \Fp(t-t') \d{t'},
   \end{aligned}
   \label{Feq}
\end{equation}
where the kernel $\kR(t) = \partial_{t}\ensav{\rn_{t}}$ is the time-dependent density of resets\footnote{%
Details of calculations throughout this paper are found in the supplemental material.}.
The integrand represents the contribution from particles that were
reset at time $t'$ and have survived until time $t$.
In the Laplace domain, $\kR(t)$ takes the simple form
$\lt \kR(\lv) = [\lv \lt \pTD(\lv)]^{-1} - 1$.
Throughout, we denote Laplace-transformed quantities by a tilde,
and the Laplace transform by $\mathcal{L}_{\lv}$.
The solution to~\eqref{Feq}
is obtained immediately as
\begin{equation}
  \lt \Fr(\lv)
  = \lt \Fp(\lv) / [\lv \lt \pTD(\lv)]
 \label{Fresetsoln}
\end{equation}

We present a number of general results on stationary and
asymptotic forms of $\Fr(t)$ that we
employ in the remainder of this paper.
Using the rule
$\lim_{t\to\infty}f(t)=\lim_{\lv\to 0}\lv\lt f(\lv)$,
the stationary average $\Fst \equiv \lim_{t\to\infty} \Fr(t)$, if it exists, is
\begin{equation}
  \Fst = \lim_{\lv\to 0} \,  \lt \Fp(\lv) / \lt \pTD(\lv)
  =  \ensav{\TD}^{-1} \int_{0}^{\infty}\pTD(t)F(t)\d{t}
 \label{Fresetstat}
\end{equation}
where the last equality holds if $\ensav{\TD}$ exists.
The following theorems
comparing the
stationary average with reset $\Fst$,
and the stationary average without reset
$\Fs \equiv \lim_{t\to\infty} F(t)$ follow directly from~\eqref{Fresetstat}~\cite{Note1}.
\begin{thmmain}
  If $\lt \Fp(\lv) \sim \lt \pTD(\lv) q(0) \text{ as } \lv\to 0$,
  where $q(\lv)\equiv \lt\Fp(\lv)/\lt \pTD(\lv)$,
  then $\Fst$ and $\Fs$ exist and $q(0) = \Fst = \Fs$.
\label{thm:stationaryone}
\end{thmmain}
\begin{thmmain}
  Suppose $\ensav{\TD}= \infty$. If $\Fs$ does not exist then $\Fst$ does not exist.
  If $\Fs$ exists, then $\Fst$ exists and $\Fst = \Fs$.
\label{thm:stationarytwo}
\end{thmmain}
A counter example is restart at constant rate $r$,
for which $\Fs = \lim_{\lv\to 0} \lv \lt F(\lv)$,
while
$\Fst = r\lt F(r)$,
which approaches $\Fs$
only as $r\to 0$.

Consider intermittent (bursty) reset, characterized by $\pTD(t)\sim (r_{0}t)^{-\nu}$.
Furthermore, assume $F(t)\sim K_{\beta}t^{\beta}$ with $\beta>0$.
Then, if $\rex > \bex + 1$ and $\rex > 1$,
the last equality in~\eqref{Fresetstat} applies.
Expansion of~\eqref{Fresetsoln} in powers of $\lv$ yields the asymptotic expressions~\cite{Note1}
\begin{equation}
  \begin{aligned}
  \Fr(t)  &\sim \frac{r_{0}^{-\rex} K_{\bex}t^{1 - \rex + \bex}}
  {(1 + \bex - \rex) \ensav{\TD}} \,
  \text{ for } 1 < \rex < \bex + 1  \\[5 pt]
  \Fr(t) & \sim \frac{\Gamma(1 + \bex -\rex)}{\Gamma(1 + \rex)\Gamma(1 + \bex)}
  K_{\bex}t^{\bex} \text{ for }
  \rex < 1.  
  \end{aligned}
  \label{bothasymp}
\end{equation}
%
%
%
Below, we apply these general results to moments of diffusion processes.

\myhead{Correlations under reset}%
An example of a functional depending on the entire most recent reset period is
\begin{equation}
  \gf(\Xt(t'), t) = \int_{0}^{t} \cos(\tshift \fq) \Xt(t)\Xt(t-\tshift) \d{\tshift},
\end{equation}
which gives the power spectral density~\cite{Krapf2019} as
$S_{T}(\fq) = 2\int_{0}^{T}\d{t}\ensav{\gf(\Xt, t)}/T$.
The power spectral density under reset is then computed from
$\ensav{\gf_{\text{r}}(\Xt, t)}$ and~\eqref{Fresetsoln}.
For Brownian motion, $\ensav{\gf(\Xt, t)}=2\mathcal{D} \fq^{-2}[1 - \cos(\fq t)]$.
Taking the power-law survival time $\pTD(t) = [\max(t,1/r_{0})r_{0}]^{-\rex}$,
we find
that the long-time, low-frequency limit changes qualitatively with $\rex$~\cite{Note1}.
In particular, for $\rex>3$
%
  $\lim_{\fq \to 0 }S_{\text{r},\infty}(\fq) = 4\mathcal{D} r_{0}^{-2}(\rex-1)/6(\rex-3)$.
For $1<\rex<3$,
 $S_{\text{r},\infty}(\fq) \sim \rex^{-1}\fq^{-(3-\rex)}  4\mathcal{D} \, \Gamma(2-\rex)\sin(\rex\pi/2) r_{0}^{1-\rex}$,
as $\fq\to 0$.
Letting $\nu\to 1$ (and thus $\ensav{\TD}\to\infty$) in the last expression we
recover the no-reset result $S_{\text{r},\infty}(\fq) \sim 4\mathcal{D} /\fq^{2}$.
It would be convenient to apply our framework to the functional $\int_{0}^{T}\d{t}\gf(\Xt, t)$,
but this is not possible because this functional includes $\Xt(t')$ conditioned on survival
not at one time, but at all times from $0$ to $T$.

\myhead{Density}
%
For $f(\Xt, t) = \delta[\xc-\Xt(t)]$,
the general results apply pointwise in $\xc$.
Eq.~\eqref{Fpdef} reduces to
$p(\xc,t)\equiv \ensav{\delta[\xc-\Xt(t)]\Ind{t<\TD}}$,
the density under decay.
Defining
$c(\xc,t)\equiv \ensav{\delta[\xc-\Xt(t)]\mid t<\TD}$,
the density under restart $\resdens(\xc,t)$
is obtained from~\eqref{Fresetsoln} as
\begin{equation}
  \lt \resdens(\xc,\lv)
  = \latrv{\lv}{\pTD(t)c(\xc,t)}/(\lv \lt \pTD(\lv)).
 \label{restartsoln}
\end{equation}
We denote the stationary density without reset by
$\cst(\xc)\equiv \lim_{t\to\infty} c(\xc,t)$
and with reset by $\prst(\xc) \equiv \lim_{t\to\infty} \resdens(\xc,t)$.
Making the substitutions
$\Fp(t) \rightarrow p(\xc,t)$,
$q(s) \rightarrow q(\xc,s)$,
$\Fst\to \prst(\xc)$,
and
$\Fs\to c(\xc)$, Theorems~\ref{thm:stationaryone}--\ref{thm:stationarytwo} apply
for the equality of stationary densities $\cst(\xc) = \prst(\xc)$.
%

\myhead{\CCFPS}
Consider the continuous time random walk (CTRW)~\cite{Montroll1965,Metzler2004,Klafter2011,Metzler2014}
in which the time of the $n$th step $T_{n}$, and $n$th position $\Wp_{n}\in \mathbb{R}$
are given by
$\Wp_{n+1} = \Wp_{n} + \xi_{n+1}$ and
$T_{n+1} = T_{n} + \tau_{n+1}$,
%
%
for $n=0,1,\dots$, and $T_{0}=0$, and $c_{0}(x)\equiv \ensav{\delta(\Wp_{0}-x)}$.
Here, $\xi_{n}$ is the random displacement,
and $\tau_{n}$ the random waiting time,
of the $n$th step.
Furthermore, $\ensav{\xi}=0$ and $\ensav{\xi^2}=\lm^2$,
where $\lm$ is the microscopic length scale.
We add to this CTRW
a single-step restart time $\Dt_{n}$ associated with
the $n$th step.
If $\Dt_{n} < \tau_{n}$, the particle takes no step,
but instead is reset according to its initial distribution $c_{0}(x)$
and the clock advances by $\Dt_{n}$ rather than $\tau_{n}$.
If instead $\tau_{n} \ge \Dt_{n}$, the particle survives the attempted
reset event, takes a step, and the walk continues.
We suppose the sets $\{\tau_{n}\}, \{\xi_{n}\}$, $\{\Dt_{n}\}$ are i.i.d.
copies of $\tau$, $\xi$, $\Dt$, respectively.
The time $\TDc$ at which the particle is reset
(after zero or more interrupted attempts at reset)
takes the form of
the generic first passage time under reset~\cite{Pal2017} given by
\begin{equation}
 \TDc =
  \begin{cases}
    \Dt   & \text{ if } \Dt < \tau \\
    \tau + \TDc' & \text{ otherwise },
  \end{cases}
 \label{Rrecurs}
\end{equation}
where $\TDc'$ is an i.i.d. copy of $\TDc$.
In this context, \textit{passage} is completion of spatial reset at time $\TDc$,
and $\tau$ plays the role of a \textit{reset} time interrupting
the spatial reset process.
We define $\pDt(t) \equiv \Pr(t<\Dt)$ and $\pTDc(t) \equiv \Pr(t<\TDc)$.

\myhead{Density under disordered reset}
%
In order to compute the density of this diffusion under disordered reset,
we first find
the density under removal or decay $p(x,t)$.
Denoting the Fourier transform by a tilde, 
the corresponding density under removal is~\cite{Lapeyre2017}
\begin{equation}
  \flt p(k,\lv) =   \frac{1 - \lt \phi_{\tau \Dt}(\lv) - \lt \phi_{\Dt \tau}(\lv)}{\lv} \frac{ \ft c_{0}(k)}
    { 1 - \lt \phi_{\tau \Dt}(\lv)  \ft\psi_{\xi}(k)},
  \label{decaydens}
\end{equation}
where
$\psi_{\xi}(x)$ is the PDF of $\xi$,
$\phi_{\Dt \tau}(t) \equiv$ \mbox{$\psi_{\Dt}(t\mid \Dt \le \tau) \Pr(\Dt \le \tau)$},
and
\begin{equation}
\phi_{\tau \Dt}(t) \equiv \psi_{\tau}(t\mid \tau < \Dt) \Pr(\tau< \Dt) \label{phitDd}.
\end{equation}
%
%
Noting that $\lt \pTDc(\lv) = \flt p(0,\lv)$,
application of the general solution~\eqref{restartsoln}
to the density of
particles that have survived removal~\eqref{decaydens}
yields immediately the particle density under continual reset
\begin{equation}
  \flt \resdens(k,\lv) = \frac{1 - \lt \phi_{\tau \Dt}(\lv)}{\lv} \frac{ \ft c_{0}(k)}
    { 1 - \lt \phi_{\tau \Dt}(\lv)  \ft\psi_{\xi}(k)}.
  \label{cresgen}
\end{equation}

We focus on subdiffusion-reaction resulting from
diverging mean waiting time $\ensav{\tau}=\infty$,
by choosing PDF
$\psi_{\tau}(t) \sim \frac{\alpha}{\ts\Gamma(1-\alpha)} (t/\ts)^{-1-\alpha}$, with $0<\alpha<1$,
for $t\gg \ts$, where $\ts$ is a microscopic timescale.
In the scaling limit
$\lm\to 0$, $\ts\to 0$ such that $\Da\equiv \lm^{2}/(2\ts^{\alpha})$ remains constant,
the density~\eqref{cresgen} becomes~\cite{Note1}
\begin{equation}
  \flt p(k,\lv) = \frac{\ft c_{0}(k)\ensav{(\lv+\rho)^{\alpha}}\lv^{-1}}{\ensav{(\lv+\rho)^{\alpha}} + k^{2}\Da},
  \label{scalelim:1}
\end{equation}
where the random rate $\rho$ is defined
by $\pDt(t) = \ensav{e^{-\rho t}}$.
Manipulation of~\eqref{scalelim:1} yields the corresponding fractional Fokker-Planck equation~\cite{Note1}
\begin{align}
  & \partialt{\resdens(x,t)} - \Da
    \partialt{} \int_0^t d t' \Kda(t - t^\prime)
    \partialxx{\resdens(x,t^\prime)}
    \nonumber\\
  & = \kRc(t) c_{0}(x) -\partialt{} \int_0^t  \kRc(t^\prime) \resdens(x,t - t^\prime) \d{t^\prime},
    \label{nldr1reset}
\end{align}
where the diffusion kernel is
$\lt\Kda(\lv) = [\lv\langle (\lv + \rho)^{\alpha - 1}]^{-1}$.
Integration over $x$ shows that $\lt \kRc(\lv) = [\lt \pTDc(\lv)\lv]^{-1} - 1$
is a special case of the reset kernel in~\eqref{Feq}.
Likewise,~\eqref{decaydens} leads to the evolution equation in the presence of decay, but not reset~\cite{Lapeyre2017}
\begin{align}
  & \partialt{p(x,t)} - \Da
    \partialt{} \int_0^t d t' \Kda(t - t^\prime)
    \partialxx{p(x,t^\prime)}
    \nonumber\\
  & = -\partialt{} \int_0^t  \kRc(t^\prime) p(x,t - t^\prime) \d{t^\prime}.
    \label{nldr1decay}
\end{align}

\myhead{Uncoupled restart}
%
The significance of~\eqref{nldr1reset} is best illustrated by comparing with
uncoupled restart.
We begin with two observations.
(\textit{i}) Let $\mathcal{O}[p(x,t)]=0$ be the evolution equation for the density $p(x,t)$
of a process under decay. Then $\mathcal{O}[p_{\text{r}}(x,t)]=h(t)c_{0}(x)$
is the evolution equation for the density under instantaneous reset
to $c_{0}(x)$,
where $h(t)$ is the function such that $p_{\text{r}}(x,t)$ is conserved.
This observation leads, for example, from~\eqref{nldr1decay} to~\eqref{nldr1reset}.
(\textit{ii}) Let $p(x,t) = w(x,t) c(x,t)$ be the density of a process under decay,
where the conserved density $c(x,t)$ satisfies the evolution equation $\mathcal{O}'[c(x,t)]=0$.
Then $\mathcal{O}'[p(x,t)/w(x,t)]=0$ yields an evolution equation for $p(x,t)$.
In particular, if $c(x,t)$ is the density of CTRW~\cite{Metzler2004},
and $\resdens(x,t)$ is the density under reset with a generic survival probability $\pTD(t)$,
then (\textit{i}) and (\textit{ii}) yield
\begin{align}
  \partialt{\resdens(x,t)}
  - &\Da \pTD(t) \partialt{} \int_{0}^{t}(t-t')^{\alpha-1}
     \partialxx{}\left[\frac{\resdens(x,t^\prime)}{\pTD(t^{\prime})}\right]\d{t'}
    \nonumber\\
   = & - \partialt{\ln \pTD(t)}
     \left[\resdens(x,t) - c_{0}(x)\right].
    \label{ffkuncoupled}
\end{align}
Eq.~\eqref{ffkuncoupled} is typical for reset processes: Reset produces a conservative current
equal to the decay rate $-\partial_{t}\ln \pTD(t)$ of the process without reset.
Even with space-dependent $w(x,t)$, the decay rate is unchanged upon
adding the reset term.
This observation is in marked contrast to the effect of introducing reset
to disordered decay,
as is seen by integrating~\eqref{nldr1reset} and~\eqref{nldr1decay} over $x$.
In~\eqref{nldr1decay}, the pure-decay rate is $-\partial_{t}\ln \pTDc(t)$.
But, in~\eqref{nldr1reset}, the decay and reset rates are equal to the reset kernel $\kRc(t)$.
\begin{figure}[htb]
  \includegraphics[width=1\columnwidth]{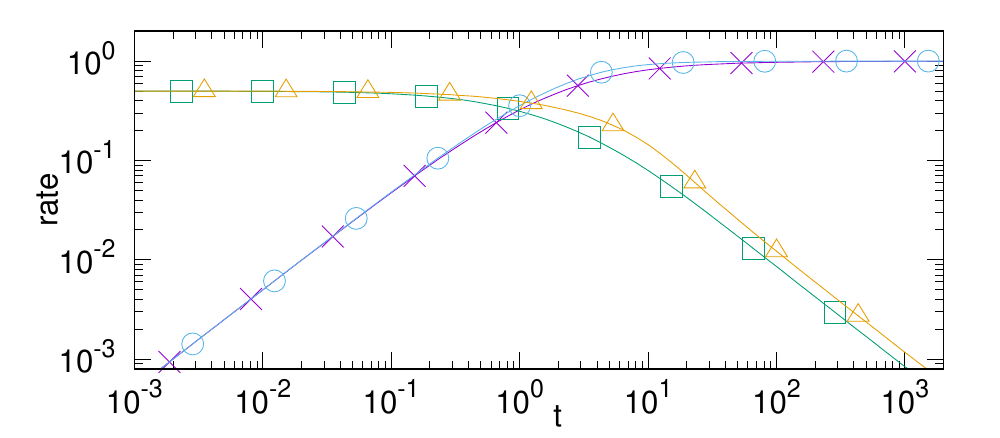}
  \caption{Decay rates $-\partial_{t}\ln \pTD(t)$ for disordered decay in~\eqref{nldr1decay}; 
    (orange triangles) $\alpha=1/2$, (green squares) $\alpha=2/3$,
    and decay/reset rate for disordered reset $\kRc(t)$ in~\eqref{nldr1reset}; 
    (purple crosses) $\alpha=1/2$,
    (blue circles) $\alpha=2/3$.
    For all curves $\pDt(t) = (1 + r_{0}t)^{-1/2}$.
  }
  \label{fig:kerndecay}
\end{figure}
As shown in~\figref{fig:kerndecay}, $-\partial_{t}\ln \pTDc(t)$
decreases monotonically, while $\kRc(t)$ increases monotonically.
This dramatic difference is due to the memory induced by coupling of transport and reset on the
microscopic level.

To highlight the effect of the dependence of $\Xt(t)$ and $\TDc$,
we consider two physical interpretations of disordered reset
that lead to different ways to uncouple the reset and transport in~\eqref{nldr1reset}.
Both yield spatially homogeneous reset described by~\eqref{ffkuncoupled}.
On one hand, since
at each step the walker samples a new random rate $\rho$ defined
by $\pDt(t)=\ensav{e^{-\rho t}}$,
we can break the dependence between
$\Xt(t)$ and reset by assuming that the walker is reset at the constant average rate $\ensav{\rho}$.
An alternative viewpoint, following~\eqref{Rrecurs}, is that
resetting with time $\Dt$ is due to the interaction of local with internal
degrees of freedom, and is itself restarted when the walker takes a step.
In this case we break the dependence by allowing the resetting process to go to completion
independently of the walker's position.
We treat the latter viewpoint first, which leads to
an ordinary CTRW with density $c(x,t)$ subject to
restart with survival time $\pDt(t)$. 
We assume the asymptotic form
$\pDt(t) \sim (r_{0}t)^{-\nu}$, with $\nu>0$,
\begin{figure}[htb]
  \includegraphics[width=1\columnwidth]{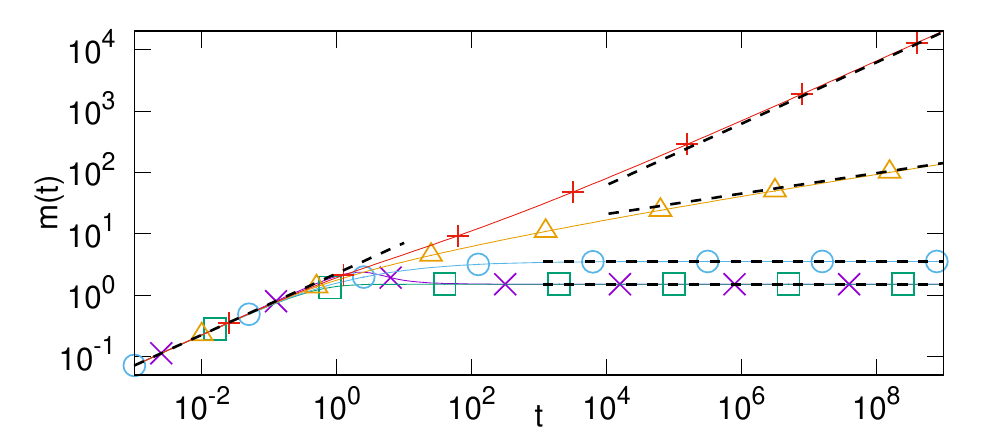}
  \caption{Mean squared displacement
    (purple crosses) under disordered decay,
    (green squares) under disordered reset,
    (blue circles) under uncoupled reset $\pDt(t)$ with $\nu>\alpha+1$,
    (orange triangles) same with $1 < \nu <\alpha+1$,
    (red pluses) same with $\nu < 1$.
    Dashed lines are exact asymptotic expansions from~\eqref{Fresetstat}, \eqref{bothasymp},
    and Ref.~\cite{Lapeyre2017}.
  }
  \label{fig:exprestart}
\end{figure}
The effect of this uncoupling is compared to the coupled process in~\figref{fig:exprestart}.
Application of Theorem~\ref{thm:stationaryone} to~\eqref{scalelim:1} shows that for all $0<\alpha<1$ and $\nu>0$,
the particle density under both coupled decay and coupled reset
tend to the same limit, $\cst(\xc) = \prst(\xc)$, and all moments exist~\cite{Note1}.
In the uncoupled, homogeneous, case,
the confining effect of reset is much weaker,
and the behavior of the moments is changed dramatically.
In this case,
application of~\eqref{Fresetstat} and \eqref{bothasymp} shows that
for $ \nu > 1 + \alpha$, the mean squared displacement (MSD)
under intermittent reset is constant at long times,
but is greater than in the disordered case.
For $1 < \nu < 1 + \alpha$, the effect of reset on the MSD is to
reduce the exponent with respect to free diffusion to the value
$1 + \alpha - \nu$.
For $\nu < 1$, the mean reset time diverges,
and the MSD grows as again as $t^{\alpha}$,
but with a reduced prefactor.

But,~\eqref{Fresetstat} and \eqref{bothasymp}
enable us to say much more about moments under intermittent reset.
Suppose
$m_{2n}(t) = \Lambda_{n} \Dgen{\alpha}^{n} t^{n\alpha}$,
is the $2n$th moment of the symmetric density of a diffusion process,
where
$\Dgen{\alpha}$ is a generalized diffusivity,
and
$\Lambda_{n}$ is a numerical factor.
For example, $\alpha=1$ for ordinary diffusion, and $0<\alpha<1$
for subdiffusion due
to fractional Brownian motion,
or a random walk on a random fractal~\cite{Stauffer1994,Havlin2002},
or
the CTRW.
It follows from~\eqref{Fresetstat} and \eqref{bothasymp},
that the $2$nth moment under reset
has a stationary limit only for $n < (\rex-1)/\alpha$.
If $\nu>1$ then moments with $n > (\rex-1)/\alpha$
grow as $t^{1-\nu+ n\alpha}$.
If $\nu < 1$, the moments grow as $t^{n\alpha}$.

In the second interpretation, we subject CTRW to decay at
constant rate $\ensav{\rho}$
finding the stationary density
$\prst(\xc)=\exp(-\lvert x \rvert/l)/(2l)$,
where $l=\sqrt{\Da \tau_{l}^{\alpha}}$,
and $\tau_{\text{u}}=1/\ensav{\rho}$~\cite{Kusmierz2018}.
We observe that replacing $\tau_{\text{u}}$ by
$\tau_{\text{c}}=\ensav{\rho^{\alpha}}^{-1/\alpha}$
gives the stationary density
under coupled reset obtained from~\eqref{scalelim:1}~\cite{Note1}.
It is easy to see that $\tau_{\text{u}}/\tau_{\text{c}} < 1$
and increases with $\alpha$ to the limit one.
Thus both ways of homogenizing the reset reduce the confinement,
although interruption of the resetting has a much stronger
effect than homogenizing rates.

\myhead{Immobilization}%
%
The solution for the density under reset~\eqref{cresgen}
is formally identical to
the Fourier-Laplace domain solution
of the standard CTRW with no decay or reset~\cite{Metzler2004},
with $\phi_{\tau \Dt}(t)$ substituted for the waiting time density $\psi_{\tau}(t)$.
We define the waiting time $\tau^{*}$
of this non-decaying CTRW
via its probability density
in accordance with~\eqref{phitDd}:
$\psi_{\tau^{*}}(t) \equiv\psi_{\tau}(t\mid \tau < \Dt) \Pr(\tau< \Dt)$.
%
%
Since $\int_{0}^{\infty} \phi_{\tau \Dt}(t)\d{t} = \Pr(\tau<\Dt)<1$,
it follows that $\psi_{\tau^{*}}(t)$ is a non-proper probability density.
In particular,
$1 - \int_{0}^{\infty}  \psi_{\tau^{*}}(t) \d{t} = \Pr(\tau^{*} = \infty) > 0$.
%
%
Thus, on the the level of the probability density,
\CFPS is equivalent to a CTRW
whose waiting time has positive probability of being infinite, immobilizing the walker.

\myhead{Conclusions and outlook}%
%
The growing importance of stochastic processes under reset
calls for a well-developed general theory.
To this end, we derived and solved in full generality equations
for averaged functionals of a stochastic process under reset
conditioned on survival in the the most recent renewal period.
We determined the condition for equality of stationary averages
with and without reset,
and asymptotic forms of generic power-law observables
under generic power-law reset times.
Applying this general framework,
we easily obtained both general and particular fundamental quantities under reset.
These include; the density, moments and spectral density.
These results subsume derivations in the literature for particular cases,
obviate the need to perform them in the future,
and sharpen our understanding by laying
clear which details are irrelevant in this context.

We applied the framework to statistically dependent
reset and diffusion arising from
diffusion under reset in a disordered medium.
The resulting generalized fractional Fokker-Planck equation
shows a spatially-dependent reset rate that differs dramatically
from removal rate in the corresponding decay model.
We compared the results with decoupled reset and diffusion,
demonstrating quantitatively enhanced confinement due to coupling.
We furthermore showed that this diffusion under disordered reset is
equivalent to an ordinary continuous time random walk
whose single-step waiting time may be infinite with positive probability.

Our general results apply not only to diffusions,
but to a wide range of applications.
An example is restarting a search process on nodes of
graph-based data structures,
which are now ubiquitous
in high-performance data analysis and machine learning~\cite{Thankachan2017,Ediger2017}.

Finally, an outstanding challenge is to extend our approach
to scenarios depending on the entire history of the of the process~\cite{Harris2017,Meylahn2015,Hollander2018,Pal2019},
or that otherwise preserve memory across reset events~\cite{Kusmierz2018,Bodrova2018}.


\vspace{5pt}

This work was supported by the European Research Council (ERC)
through the project MHetScale (Contract number 617511)



%



\newpage ~ \newpage

\setcounter{equation}{0}
\setcounter{section}{0}
\setcounter{figure}{0}
\setcounter{table}{0}
\setcounter{page}{1}
\makeatletter
\renewcommand{\theequation}{S\arabic{equation}}
\renewcommand{\thefigure}{S\arabic{figure}}
\renewcommand{\thesection}{\Roman{section}}

\onecolumngrid

\noindent
\textbf{\textsf{\Large Supplemental Information for the article ``\@title''}}

\section{General approach to  processes  under reset\label{supps:restart}}

\subsection{Average over a functional of the process under reset}

\subsubsection{Derivation of the evolution of average of functionals of $\Xt$ under reset: $\ensav{\fr\left(\Xt, t\right)}$\label{sec:resetderivation}}

As in the main text, we consider an arbitrary stochastic process $\Xt(t)$
and random death time $\TD$. The process under reset is obtained by
instantaneously resetting the particle to its initial state (or distribution over states)
every time it dies, that is, after
i.i.d. time intervals $\{\TD_{n}\}$, with $\TD_{n} \overset{d}{=} \TD$.
The time of the $n$th reset $\rt_{n}$ is then determined by
$\rt_{1}=\TD_{1}$ and $\rt_{n+1} = \rt_{n}+\TD_{n+1}$.
It follows that the number of resets performed up to time $t$ is the renewal process
\begin{equation}
\rn_{t}\equiv \max\{n: \rt_{n} \le t\}.
  \label{Ntdef:supps}
\end{equation}
Thus, at any time $t$, the
time of the most recent reset is $T_{\rn_{t}}$.
The process under reset is given by $\Xtr(t) \equiv \Xt_{\rn_{t}}(t-T_{\rn_{t}})$.
We study the average of functionals of the reset process
$\Fr(t)\equiv \ensav{f(\Xtr(t'), t)}$
that depend only on the most recent renewal period
up to time $t$.
That is,
the average depends on $\Xtr(t')$ for $T_{\rn_{t}} \le t' < T_{\rn_{t}+1}$.
We use the following simpler definition of the process under reset
that is equivalent for computing averages restricted to the most recent renewal period
\begin{equation}
  \Xtr(t) \equiv \Xt(t-T_{\rn_{t}}).
  \label{resetdef:supps}
\end{equation}
Note that~\eqref{resetdef:supps} involves only a single copy of the process.
Then
%
   $f\left(\Xt_{\text{r}}, t \right) = f\left(\Xt, t - T_{\rn_{t}} \right)$.
%
For notational convenience  we write $\ensav{\fr\left(X, t\right)}$ for $\ensav{f(\Xtr(t'), t)}$.
Partitioning the expectation by number of reset events $n$ we write
\begin{equation*}
\begin{aligned}
 \ensav{\fr\left(\Xt, t\right)}
   = &  \sum_{n=0}^{\infty} \ensav{ \Ind{N_{t}=n} f\left(\Xt, t - T_{n}\right)}
   = & \sum_{n=0}^{\infty} \ensav{ \Ind{T_{n} \le t < T_{n+1}} f\left(\Xt, t - T_{n}\right) }.
 \end{aligned}
\end{equation*}
Writing $T_{n+1}$ as $T_{n}+\TD_{n+1}$, we have
\begin{equation*}
  \begin{aligned}
 \ensav{\fr\left(\Xt, t\right)} =
 \sum_{n=0}^{\infty} \ensav{ \Ind{T_{n} \le t} \ \Ind{t < T_{n}+\TD_{n+1}} \ f\left(\Xt, t - T_{n}\right) }.
\end{aligned}
\end{equation*}
Introducing an integral and delta function to replace the random variable $T_{n}$ by the number $t'$, we write
\begin{equation*}
\begin{aligned}
  \ensav{\fr\left(\Xt, t\right)} =
   \sum_{n=0}^{\infty} \int_{0}^{\infty} \d{t'} \Big\langle \delta(t'-T_{n})  \Ind{t' \le t} \ \Ind{t - t' < \TD_{n+1}}
  f\left(\Xt, t - t'\right)
  \Big\rangle.
\end{aligned}
\end{equation*}
We replace $\Ind{t' \le t}$ by an upper limit on the integral
\begin{equation*}
\begin{aligned}
  \ensav{\fr\left(\Xt, t\right)} =
   \sum_{n=0}^{\infty} \int_{0}^{t} \d{t'} \ensav{ \delta(t'-T_{n}) \ \Ind{t - t' < \TD_{n+1}} \ f\left(\Xt, t - t'\right)}.
\end{aligned}
\end{equation*}
Observing that $T_{n}$ and $\TD_{n+1}$ are independent, we factor the expectation,
and furthermore replace $\TD_{n+1}$ by $\TD$ because they
are equal in distribution, obtaining
\begin{equation}
\begin{aligned}
  \ensav{\fr\left(\Xt, t\right)} =
 \sum_{n=0}^{\infty} \int_{0}^{t} \d{t'} \ensav{ \delta(t'-T_{n})} \ensav{\Ind{t - t' < \TD} \  f\left(\Xt, t - t'\right)}.
\end{aligned}
 \label{finalstep:supps}
\end{equation}
Defining
\begin{align}
  \psi_{n}(t) = \ensav{\delta(t-T_{n})} \quad \text{ and } \quad \kR(t) = \sum_{n=0}^{\infty} \psi_{n}(t) - \delta(t)
  \label{kandpsidef:supps}
\end{align}
we obtain our main result
\begin{equation}
   \begin{aligned}
     \ensav{\fr\left(\Xt, t\right)}
     &=  \int_{0}^{t}  [\kR(t')+\delta(t')]  \ensav{f(\Xt,t-t') \Ind{t - t' < \TD}} \d{t'} \\
     &= \ensav{f(\Xt,t) \Ind{t < \TD}} + \int_{0}^{t} \kR(t - t') \ensav{f(\Xt, t') \Ind{t' < \TD}} \d{t'} \\
     &= \Fp(t) + \int_{0}^{t} \kR(t - t') \Fp(t') \d{t'},
   \end{aligned}
   \label{resdenseq:supps}
\end{equation}
where we defined $\Fp(t)\equiv \ensav{f(\Xt,t) \Ind{t < \TD}}$.
If $\Xt(t)$ and $R$ are independent, then~\eqref{resdenseq:supps} reduces to
\begin{equation}
   \begin{aligned}
     \ensav{\fr\left(\Xt, t\right)}
     &= \pTD(t) \ensav{f(\Xt,t)} + \int_{0}^{t} \kR(t - t') \ \pTD(t') \ensav{f(\Xt, t')} \d{t'},
   \end{aligned}
   \label{resdenseqind:supps}
 \end{equation}
where $\pTD(t)\equiv \Pr(t<\TD)$.
Choosing the constant function $f(\Xt,t)\equiv 1$ and taking the Laplace transform,
we find the expression for the kernel in the Laplace domain,
\begin{equation}
  \lt \kR(\lv)=\frac{1}{\lv \lt \pTD(\lv)} - 1.
  \label{kern:supps}
\end{equation}
As in the main text, we use the notation
$F(t) \equiv \ensav{f(\Xt,t) \mid t < \TD}$
and
$\Fr(t) \equiv \ensav{f(\Xt,t-T_{\rn_{t}})}$
to write~\eqref{resdenseq:supps} as
\begin{equation}
   \begin{aligned}
     \Fr(t)
     = \pTD(t) F(t)
     +  \int_{0}^{t} \kR(t - t') \ \pTD(t') F(t')\d{t'},
   \end{aligned}
   \label{Feq:supps}
\end{equation}
with solution
\begin{equation}
  \lt \Fr(\lv) = \frac{\lt \Fp(\lv)}{\lv \lt \pTD(\lv)}
  = \frac{\latrv{\lv}{\pTD(t)F(t)}}{\lv \lt \pTD(\lv)}.
 \label{Fresetsoln:supps}
\end{equation}
Employing the identity $\lim_{t\to\infty} f(t) = \lim_{\lv\to 0} \lv \lt f(\lv)$,
we find that the stationary solution,
if it exists, is
\begin{equation}
  \Fst \equiv \lim_{t\to\infty} \Fr(t) =
  \lim_{\lv\to 0}  \frac{\lt \Fp(\lv)}{\lt \pTD(\lv)}
  = \frac{\int_{0}^{\infty} \pTD(t)F(t)\d{t}}{\ensav{\TD}},
 \label{Fresetstat:supps}
\end{equation}
where the last equality holds if $\ensav{\TD}$ exists.
Defining $\Fs \equiv \lim_{t\to\infty} F(t)$ and
$q(\lv)\equiv \lt \Fp(\lv) / \lt \pTD(\lv)$,
we consider the condition
\begin{equation}
  \lt \Fp(\lv) = \latrv{\lv}{\pTD(t)F(t)} \sim \lt \pTD(\lv) q(0) \text{ as }  \lv\to 0,
  \label{eqcond:supps}
\end{equation}
where $q(0) \equiv \lim_{\lv\to 0}q(\lv)$.
We have the following theorems relating the stationary average with reset $\Fst$ and without reset $\Fs$.
\begin{thm}
  If $\lt \Fp(\lv) \sim \lt \pTD(\lv) q(0) \text{ as } \lv\to 0$,
  where $q(\lv)\equiv \lt\Fp(\lv)/\lt \pTD(\lv)$,
  then $\Fst$ and $\Fs$ exist and $q(0) = \Fst = \Fs$.
\label{thm:stationaryone:supps}
\end{thm}
\begin{thm}
  Suppose $\ensav{\TD}= \infty$. If $\Fs$ does not exist then $\Fst$ does not exist.
  If $\Fs$ exists, then $\Fst$ exists and $\Fst = \Fs$.
\label{thm:stationarytwo:supps}
\end{thm}
These theorems are demonstrated as follows.
If condition ~\eqref{eqcond:supps} holds,
the limit~\eqref{Fresetstat:supps} immediately gives that $\Fst$ exists and $\Fst = q(0)$.
But,~\eqref{eqcond:supps} also implies $\Fp(t) \sim \pTD(t) q(0)$ as $t\to\infty$,
which in turn implies that $\Fs \equiv \lim_{t\to\infty} F(t) = q(0) = \Fst$.
Thus~\eqref{eqcond:supps} implies that the stationary averages with and without reset both exist
and are equal: $\Fst = \Fs$.
Suppose instead that $\Fst$ exists and that
$\ensav{\TD}=\lim_{\lv\to 0} \lt \pTD(s) = \infty$.
Then~\eqref{eqcond:supps} must hold in order
for the divergences in the numerator and denominator in~\eqref{Fresetstat:supps}
to cancel, and once again $\Fst = \Fs$.
If $\ensav{\TD}<\infty$, then~\eqref{eqcond:supps} is satisfied only if the first non-analytic term
in the expansion of $\lt \pTD(\lv)$ is of lower order than the first non-analytic term in $q(\lv)$.
The second part of theorem~\ref{thm:stationarytwo:supps} follows from the contrapositive of the first part.

We stress that all results in this section apply to arbitrary functionals $f$ and processes $\Xt$ and $R$.
A particular application will have additional degrees of freedom, such as the point $\xc$ in the state space of $\Xt$
for the occupation density $c(\xc, t)$,
or the time shift $\tshift$ in the autocorrelation function $C(t;\tshift)$. These are carried along as parameters
when applying the results of this section.
Below we show that the  application is often quite straightforward, reproduces known results, and produces new ones.

\subsubsection{Reset at constant rate}

For restart at constant rate $r$, we have $\pTD(t)=e^{-rt}$.
Then~\eqref{kern:supps} gives $\kR(t) = r$.
According to~\eqref{Feq:supps}
the averaged function of $\Xt(t)$ under reset is then
\begin{equation}
   \begin{aligned}
     \Fr(t) = e^{-rt} F(t) + r \int_{0}^{t} e^{-r t'}  F(t') \d{t'},
   \end{aligned}
   \label{resdensconst:supps}
\end{equation}
with solution given by~\eqref{Fresetsoln:supps}
\begin{equation}
  \lt \Fr(\lv) = \lt F(\lv + r) \frac{\lv + r}{\lv},
 \label{Fconstsoln:supps}
\end{equation}
and stationary solution given by~\eqref{Fresetstat:supps}
\begin{equation}
  \lim_{t\to\infty} \Fr(t) =  r \lt F(r).
  \label{Fresetconststat:supps}
\end{equation}

\subsection{Specific functionals  $f(\Xt, t)$ under reset}

A main point of these examples is to demonstrate that
for $f(\Xt, t)$ satisfying the conditions given in~\secref{sec:resetderivation}
obtaining an expression for its average under reset $\ensav{\fr(\Xt, t)}$,
and attendant properties,
requires no further probabilistic reasoning.
Analysis of the particular problem is then reduced to computing $\ensav{f(\Xt, t)|t<\TD}$.

\subsubsection{Occupation density}

Here we compute the particle (or occupation) density,
usually the first object of interest when studying a process under reset.
Taking
\begin{equation}
  f(\Xt, t) = \delta[\xc-\Xt(t)],
\end{equation}
Eq.~\eqref{resdenseq:supps} immediately gives us
\begin{equation}
  \resdens(\xc,t)  = \pTD(t) c(\xc,t) + \int_{0}^{t} \kR(t') \pTD(t-t') c(\xc,t-t') dt',
 \label{convonea:supps}
\end{equation}
where we denote the particle density under reset by
\begin{equation}
  \resdens(\xc,t) \equiv \ensav{\delta[\Xt_{\text{r}}(t)-\xc]},
 \label{resdens:supps}
\end{equation}
and define
\begin{equation}
  c(\xc,t) \equiv \ensav{\delta[\Xt(t)-\xc] \mid t < \TD}.
 \label{cdef:supps}
\end{equation}
Eq.~\eqref{convonea:supps}
has been derived in many contexts, including:
Eq.(2) in Ref.~\cite{Meylahn2015} in the context of a one-dimensional continuous time
Markov process; Eq.(5) in Ref.~\cite{Evans2014} in the context of diffusion in three dimensions;
Eq.(2) in Ref.~\cite{Eule2016}
in the context of a motion governed by a Langevin equation under non-Poissonian reset;
Eq.(10) in Ref.~\cite{Economou2003}
and on the last page of Ref.~\cite{Kyriakidis1994} in the context of population dynamics.
Note that Theorem~\ref{thm:stationarytwo:supps} applies to densities.
In particular, if $c(\xc,t)$ has no stationary density,
then $\ensav{\TD}=\infty$ implies no stationary density under reset.

\subsubsection{Intermittent (power-law) reset of power-law averages \label{sec:moments}}

Suppose that the averaged functional $F(t)$ varies asymptotically as a
power
$F(t) \sim K_{\bex}t^{\bex}$,
and that the survival probability
varies as
$\pDt(t) \sim (r_{0}t)^{-\rex}$.
Then
$\Fp(t)\sim r_{0}^{-\rex} K_{\bex} t^{\bex-\rex}$.
In the following, we compute asymptotic forms of $\Fr(t)$ via~\eqref{Fresetsoln:supps} and~\eqref{Fresetstat:supps}.
Suppose that $\rex > \bex + 1$ and $\rex > 1$.
Then both
$\ensav{\Dt} = \int_{0}^{\infty}\pDt(t)\d{t} = \lim_{\lv\to 0} \lt \pDt(\lv)$
and
$\lt \Fp(0) \equiv \lim_{\lv\to 0} \lt \Fp(\lv)$
exist,
and $\Fr(t)$ has the stationary limit
given directly by~\eqref{Fresetstat:supps},
\begin{equation}
  \Fst = \ensav{\Dt}^{-1}\int_{0}^{\infty}\pDt(t)F(t) \d{t}.
  \label{stationaryrho:supps}
\end{equation}
Now suppose $1 < \rex < \bex + 1$ so that $\ensav{\Dt}$ exists,
but $\lt \Fp(0)$ does not.
Then,
$\lt \Fp(\lv) \sim r_{0}^{-\rex} K_{\bex} \Gamma(1+\bex-\rex)\lv^{\rex - \bex - 1}$
as $\lv\to 0$.
And $\Fr(t)$ grows for large $t$ as
\begin{equation}
  \Fr(t) \sim \frac{r_{0}^{-\rex} K_{\bex}}
  {(1 + \bex - \rex) \ensav{\Dt}} \, t^{1 - \rex + \bex}.
    \label{alteredpower:supps}
\end{equation}
Finally, suppose $\rex < 1$, and $\rex < \bex + 1$,
so that neither $\lt \Fp(0)$ nor $\ensav{\Dt}$ exist.
Then $\lt \pDt(\lv) \sim r_{0}^{-\rex} \lv^{\rex-1}$ as $s\to 0$,
and for large $t$ we find
\begin{equation}
  \Fr(t) \sim \frac{\Gamma(1 + \bex -\rex)}{\Gamma(1 + \rex)\Gamma(1 + \bex)}
  K_{\bex}t^{\bex}.
  \label{norhomean:supps}
\end{equation}

In the main text, we consider the concrete choice $\pDt(t) = \ensav{e^{-t\rho}} = (1 + r_{0}t)^{-\nu}$,
with corresponding PDF $\psi_{\rho}(r) = r_{0}^{-\nu}r^{\nu-1}\exp(-r/r0)/\Gamma(\nu)$.
Then $\ensav{\Dt}=[r_{0}(\nu-1)]^{-1}$.
Note that~\eqref{norhomean:supps} is independent of details of $\pDt(t)$.
Eq.~\eqref{alteredpower:supps}, which depends on $F(t)$ only via its asymptotic form, becomes
\begin{equation}
  \Fr(t) \sim \frac{(\nu-1)r_{0}^{1-\rex} K_{\bex}}
  {(1 + \bex - \rex)} \, t^{1 - \rex + \bex}, \ \text{ for } 1 < \rex < \bex + 1
    \label{alteredpowergamma:supps}
\end{equation}
Eq.~\eqref{stationaryrho:supps} depends on both $F(t)$ and $\pDt(t)$ for all $t\ge 0$.
So, we choose $F(t) = K_{\bex}t^{\bex}$
(which is more restrictive than $F(t) \sim K_{\bex}t^{\bex}$),
and find
\begin{equation}
  \Fst = \frac{\Gamma(1+\bex)\Gamma(1+\bex-\rex)}{\Gamma(\nu-1)} K_{\bex}r_{0}^{-\nu},
   \ \text{ for }  \rex > \bex + 1, \ \rex > 1.
\end{equation}

We also consider the Pareto distribution with survival probability
\begin{equation}
  \pDt(t) = [\max(t,t_{0})/t_{0}]^{-\rex}, \text{ for } \rex>0
  \label{paretosurv:supps}
\end{equation}
and its Laplace transform
\begin{equation}
  \lt \pDt(\lv) = \lv^{-1} \left[ 1 - e^{-\lv t_{0}} + (\lv t_{0})^{\rex} \Gamma(1-\rex, \lv t_{0}) \right],
  \label{ltsurvprobpower}
\end{equation}
where $\Gamma(n,z)$ is the upper incomplete gamma function.
The mean survival time $\ensav{\Dt}$ is
\begin{equation}
  \ensav{\Dt} = \lim_{s\to0} \lt \pDt(\lv) = t_{0} \rex/(\rex-1) \ \text{ for } \rex > 1,
  \label{paretosurvmean:supps}
\end{equation}
and does not exist for $\rex \ge 1$.

\subsubsection{Power spectral density  \label{sec:PSD}}

Suppose
\begin{equation}
  \Xt(t)\in\mathbb{R}, \ \ensav{\Xt(t)}=0, \ \text{ and } \ \Xt(t) = 0 \text{ for } t \le 0.
  \label{Xtposdef:supps}
\end{equation}
The power spectral density (PSD) is defined by
\begin{equation}
  S_{T}(\fq) = \frac{1}{T} \int_{0}^{T} \d{t_{2}} \int_{0}^{T} \d{t_{1}}
  \cos([t_{1} - t_{2}]\fq)
  \ensav{\Xt(t_{1}) \Xt(t_{2})},
  \label{PSD:supps}
\end{equation}
which we rewrite as
\begin{equation}
  S_{T}(\fq) = \frac{2}{T} \int_{0}^{T}  \ensav{\gf(\Xt, t)} \d{t},
\end{equation}
where the functional $g$ is
\begin{equation}
  \gf(\Xt, t) \equiv \int_{0}^{t} \d{\tshift} \cos(\tshift \fq) \Xt(t)\Xt(t-\tshift),
\end{equation}
and $\ensav{\gf(\Xt, t)}$ is written
\begin{equation}
  \ensav{\gf(\Xt, t)} = \int_{0}^{t} \d{\tshift} \cos(\tshift \fq) C_{0}(t;\tshift),
  \label{gfavg:supps}
\end{equation}
and the autocorrelation function $C_{0}(t;\tshift)$ is given by
\begin{equation}
  C_{0}(t;\tshift) \equiv \ensav{\Xt(t)\Xt(t-\tshift)}, \quad \tshift > 0.
  \label{autocorrelation0:supps}
\end{equation}
Using~\eqref{resdenseqind:supps}, we find that the
power spectral density under reset $S_{\text{r},T}(\fq)$
for arbitrary random reset time is
\begin{equation}
  S_{\text{r},T}(\fq) = \frac{2}{T} \int_{0}^{T}  \ensav{\gf_{\text{r}}(\Xt, t)} \d{t},
  \label{PSDrgen:supps}
\end{equation}
where $\gf_{\text{r}}(\Xt, t)$ is $\gf(\Xt, t)$ under reset.
Using~\eqref{Fresetsoln:supps},~\eqref{PSDrgen:supps},
the rule $\lt h(\lv)/\lv = \lt H(\lv) - H(0)/\lv$,
where $H(t)=\int h(t) \d{t}$,
the solution is
\begin{equation}
  S_{\text{r},T}(\fq) = \frac{2}{T}
   \ilatrv{T}{\frac{\latrv{\lv}{p(t)\ensav{\gf(\Xt, t)}}}{\lv^{2} \lt p(\lv)}}.
\end{equation}
Assuming restart at constant rate $r$, defining $F(t) = \ensav{\gf(\Xt, t)}$, and using~\eqref{Fconstsoln:supps},
the Laplace-domain solution is given by
\begin{equation}
  S_{\text{r},T}(\fq)  = \frac{2}{T} \ilatrv{T}{\lt F(\lv + r) \frac{\lv + r}{\lv^{2}}}.
  \label{PSDrconst:supps}
\end{equation}
For Brownian motion the autocorrelation function~\eqref{autocorrelation0:supps} is given by
\begin{equation}
  C_{0}(t;\tshift) = 2\mathcal{D} (t-\tshift) u(t-\tshift),
  \label{autocorrelationbrownian:supps}
\end{equation}
where $u(t)$ is the unit step function, and $\mathcal{D}$ the diffusion coefficient.
Then~\eqref{gfavg:supps} becomes
\begin{equation}
  F(t) = \ensav{\gf(\Xt, t)} = 2\mathcal{D} \fq^{-2}[1 - \cos(\fq t)],
  \label{browniang:supps}
\end{equation}
with Laplace transform $\lt F(\lv) = 2\mathcal{D}/[\lv(\fq^{2} + \lv^{2})]$.
The Laplace transformed quantity in~\eqref{PSDrconst:supps} is then $[1/(\lv^2  (\fq^2 + (\lv+r)^2))$,
which is easily inverted to find that the PSD for Brownian motion under constant reset rate $r$ is
\begin{equation}
  S_{\text{r},T}(\fq) = \frac{4\mathcal{D}}{\fq^{2}+r^{2}}
  \left[1 - \frac{1}{\fq(\fq^{2} + r^{2})T} (2\fq r - 2\fq r \cos(\fq T)e^{-rT} -
    (r^{2} - \fq^{2})\sin(\fq T) e^{-rT})\right]
\end{equation}
This expression appears as Eq.(7) in Ref.~\cite{Majumdar2018} where various limits are discussed
and further application to fractional Brownian motion is presented.

Returning to general $\TD$ and $\Xt(t)$,
if the limit $T\to\infty$ in~\eqref{PSDrgen:supps} exists,
then applying Theorem~\ref{thm:stationarytwo:supps},
we have
\begin{equation}
  \lim_{T\to\infty} S_{\text{r},T}(\fq) =
   \frac{2 \lim_{\lv\to 0}\latrv{\lv}{p(t)\ensav{\gf(\Xt, t)}}}{\ensav{\TD}}.
\end{equation}
For Brownian motion,~\eqref{browniang:supps} gives
\begin{equation}
  \lim_{T\to\infty} S_{\text{r},T}(\fq) =
   \frac{4\mathcal{D}}{\fq^{2}}
   \frac{\lim_{\lv\to 0}\latrv{\lv}{p(t)[1-\cos(\fq t)]}}{\ensav{\TD}}.
   \label{brownianPSDstat:supps}
\end{equation}
Choosing for $p(t)$ the Pareto survival time~\eqref{paretosurv:supps}, we have
\begin{equation}
  \latrv{\lv}{p(t)[1-\cos(\fq t)]} =
  \frac{e^{-\lv t_{0}}\left[\fq^{2}(e^{\lv t_{0}} - 1) - \lv^{2}[1-\cos(\fq t_{0})] - \fq\lv \sin(\fq t_{0}) \right]}{\lv(\lv^{2}+\fq^{2})}
   + t_{0} \int_{1}^{\infty}x^{-\rex} e^{-\lv t_{0} x}[1-\cos(\fq t_{0}x)]\d{x}.
 \label{stationaryPSDbrownian:supps}
\end{equation}
Using~\eqref{paretosurvmean:supps}, the stationary PSD for Brownian motion~\eqref{brownianPSDstat:supps}
under Pareto-distributed reset
becomes
\begin{equation}
  S_{\text{r},\infty}(\fq) \equiv \lim_{T\to\infty} S_{\text{r},T}(\fq) =
   \frac{4\mathcal{D}}{\fq^{2}}
    \left\{ 1 - \frac{\rex-1}{\rex}\left[\int_{1}^{\infty}x^{-\rex} \cos(\fq t_{0}x) \d{x}
        +\frac{\sin(\fq t_{0})}{\fq t_{0}}\right] \right\}
    \label{brownparetoPSD:supps}
\end{equation}
The limiting cases are instructive.
The low frequency limit is, for $\rex>3$,
\begin{equation}
  \lim_{\fq \to 0 }S_{\text{r},\infty}(\fq) = \frac{4\mathcal{D} t_{0}^{2}(\rex-1)}{6(\rex-3)},
\end{equation}
and for $\rex<3$,
\begin{equation}
  S_{\text{r},\infty}(\fq) \sim \fq^{-(3-\rex)}  \frac{4\mathcal{D} \, \Gamma(2-\rex)\sin(\rex\pi/2) t_{0}^{\rex-1}}{\rex}
  \ \text{ as } \fq \to 0.
  \label{brownparetoPSDlowfreq:supps}
\end{equation}
This means that for $\rex<3$,
the effect of resetting is not strong enough to cut off low frequency correlations.
As $\rex\to 1$ in~\eqref{brownparetoPSDlowfreq:supps}, the PSD approaches the
stationary value for Brownian motion without reset $4\mathcal{D}/\fq^{2}$, which is expected since
the mean reset time diverges at $\rex=1$.
We also recover the result for Brownian motion for large minimum reset time $t_{0}$,
$\lim_{t_{0}\to\infty} S_{\text{r},\infty}(\fq) = 4\mathcal{D}/\fq^{2}$.
The high frequency limit $S_{\text{r},\infty}(\fq) \sim 4\mathcal{D}/\fq^{2}$ as
$\fq\to\infty$ also approaches that for Brownian motion.
The limit of rapid resets gives $\lim_{t_{0}\to\infty}S_{\text{r},\infty}(\fq)=0$,
as does reset at constant rate with $r\to\infty$~\cite{Majumdar2018}.
Finally, as $\beta\to\infty$, reset occurs at the non-random time $t_{0}$ and correlations
are destroyed.
This is equivalent to computing the PSD~\eqref{PSD:supps} at time $T=t_{0}$.
As expected,~\eqref{brownparetoPSD:supps} gives the PSD for Brownian motion computed at time $t_{0}$
\begin{equation}
  \lim_{\rex \to\infty} S_{\text{r},\infty}(\fq) = \frac{4\mathcal{D}}{\fq^{2}}\left[1-\frac{\sin(\fq t_{0})}{\fq t_{0}}\right].
\end{equation}

\subsubsection{Autocorrelation function \label{sec:autocorrelation}}

Assuming~\eqref{Xtposdef:supps}, the functional $f(\Xt, t) = \Xt(t)\Xt(t-\tshift)$,
and using~\eqref{autocorrelation0:supps},
the autocorrelation function under reset
$ C_{r}(t;\tshift) \equiv \ensav{\Xt_{\text{r}}(t)\Xt_{\text{r}}(t-\tshift)}$
is obtained directly from~\eqref{resdenseqind:supps} as
\begin{equation}
  C_{r}(t;\tshift) = p(t) C_{0}(t;\tshift) + \int_{0}^{t} \kR(t') p(t-t') C_{0}(t-t';\tshift) \d{t'}.
  \label{autocorrelation:supps}
\end{equation}
Eq.~\eqref{autocorrelation:supps} for the particular case of constant reset rate,
that is $p(t)=e^{-rt}$,
appears as Eq.(4) in~\cite{Majumdar2018}.
Now let us assume that $\Xt(t)$ is Brownian motion.
The Laplace transform of the autocorrelation function for Brownian motion~\eqref{autocorrelationbrownian:supps} is
\begin{equation}
  \lt C_{0}(\lv;\tshift) = 2\mathcal{D} \frac{e^{-\tshift \lv}}{\lv^{2}}.
\end{equation}
We furthermore assume constant reset rate $r$. Then the result for general functional~\eqref{Fresetconststat:supps} gives
the stationary (fixed $\tshift$, large $t$) solution
%
  $\lim_{t\to\infty} C_{r}(t;\tshift) = r \lt C_{0}(r;\tshift) = \frac{2\mathcal{D}}{r} e^{-\tshift r}$.
%
Eq.~\eqref{Fconstsoln:supps} gives the Laplace-domain solution for finite $t$
\begin{equation}
  \lt C_{r}(\lv;\tshift) = 2\mathcal{D}\frac{e^{-\tshift(\lv + r)}}{\lv(\lv + r)},
\end{equation}
which upon inversion gives the autocorrelation function for Brownian motion under constant reset rate
\begin{equation}
  C_{r}(t;\tshift) = \frac{2\mathcal{D}}{r} \left(e^{-\tshift r} - e^{- t r}\right) u(t-\tshift),
\end{equation}
in agreement with Eq.(5) in Ref.~\cite{Majumdar2018}.
By Theorem~\ref{thm:stationarytwo:supps}, $\lim_{t\to\infty} C_{r}(t;\tshift)$
exists only for finite $\ensav{\TD}$.
For power-law survival time~\eqref{paretosurv:supps}
with corresponding mean~\eqref{paretosurvmean:supps},
the stationary value
of the autocorrelation function of Brownian motion~\eqref{autocorrelationbrownian:supps}
under reset is given via~\eqref{Fresetstat:supps} as
\begin{equation}
 \lim_{t\to\infty} C_{r}(t;\tshift) = \frac{2\mathcal{D} t_{0}^{\rex-1}}{\rex(\rex-2)}  \tshift^{2-\rex}, \ \text{ for } \rex > 2, \ \tshift > t_{0}.
\end{equation}
%

\subsection{Statistical meaning of the kernel $\kR(t)$}

Here show that $\kR(t)$ is the density of the mean number or resets, ie,
$\kR(t) = \partial_{t}{\ensav{\rn_{t}}}$.
Note that $\rn_{t}$, the number of resets up to time $t$,
as defined in~\eqref{Ntdef:supps}, may also be written
\begin{equation*}
  \rn_{t} = \sum_{n=1}^{\infty} \Ind{T_{n}\le t},
\end{equation*}
so that the mean number of resets up to time $t$ is
\begin{equation*}
  \ensav{\rn_{t}} = \sum_{n=1}^{\infty} \Pr(T_{n}\le t),
\end{equation*}
and the associated density of the mean number of resets (not a probability density) is
\begin{equation*}
  \partialt{\ensav{\rn_{t}}} = \sum_{n=1}^{\infty} \psi_{n}(t),
\end{equation*}
where $\psi_{n}(t)\equiv \partial_{t} \Pr(T_{n}\le t)$.
In words, $\psi_{n}(t)$ is the PDF of the time at which the $n$th reset occurs.
It is easy to see that $\psi_{n}(t)$ is the $n$-fold convolution of $\psi_{1}(t)$~\cite{Klafter2011,Metzler2000}.
Recall that $T_{1}= \TD_{1}$, $\TD_{1} \overset{d}{=} \TD$, and $-\partial_{t} p_{\TD}(t)=\psi_{\TD}(t)$,
where the survival probability of the reset time $p_{\TD}(t) = \Pr(t<\TD)$,
and $\psi_{\TD}(t)$ is the corresponding PDF.
In the Laplace domain, we then have $\lt \psi_{n}(\lv) = \lt \psi_{\TD}^{n}(\lv)$,
and
\begin{equation*}
  \latr{\partialt{\ensav{\rn_{t}}}} = \frac{\lt\psi_{\TD}(\lv)}{1- \lt\psi_{\TD}(\lv)}.
\end{equation*}
Referring to~\eqref{kern:supps} we see that
%
  $\kR(t) = \partialt{\ensav{\rn_{t}}}$.
%

\section{Diffusion and reset on a disordered medium}

\subsection{Generalized fractional reaction diffusion equation
   for \CFPS \label{supps:genFFP}}

We begin with the generalized Montroll-Weiss equation in the scaling limit
for disordered diffusion under decay or removal~\cite{Lapeyre2017}
\begin{equation}
  \flt p(k,\lv) = \frac{\ft c_{0}(k)\ensav{(\lv+\rho)^{\alpha-1}} }
  {\ensav{(\lv+\rho)^{\alpha}} + k^2 \Da}.
   \label{genMW:supps}
\end{equation}
Recall that the spatial reset time $\TDc$ is itself given
by the first passage time $\Dt$ subject to reset time $\tau$
\begin{equation}
 \TDc =
  \begin{cases}
    \Dt   & \text{ if } \Dt < \tau \\
    \tau + \TDc' & \text{ otherwise }.
  \end{cases}
 \label{Rrecurs:supps}
\end{equation}
The Laplace transform of the survival probability $p_{\TDc}(t)$
is obtained by setting $k$ to zero in~\eqref{genMW:supps}
\begin{equation}
  \lt p_{\TDc}(\lv) = \frac{\ensav{(\lv+\rho)^{\alpha-1}} }
  {\ensav{(\lv+\rho)^{\alpha}}},
  \label{dissurv:supps}
\end{equation}
where $p_{\Dt}(t) = \ensav{e^{-\rho t}}$.
Applying the generic formula~\eqref{Fresetsoln:supps}, we obtain
the density of \CFPS in the Fourier-Laplace domain
\begin{equation}
\begin{aligned}
  \flt \resdens(k,\lv) = \frac{\lv^{-1} \ft c_{0}(k)\ensav{(\lv+\rho)^{\alpha}} }
  {\ensav{(\lv+\rho)^{\alpha}} + k^2 \Da}.
\end{aligned}
\label{ctrwfptresetdens:supps}
\end{equation}
Multiplying by the denominator, we have
\begin{equation}
\begin{aligned}
   \flt \resdens(k,\lv) \ensav{(\lv+\rho)^{\alpha}}
  = -\Da k^{2} \flt \resdens(k,\lv) + \lv^{-1} \ft c_{0}(k) \ensav{(\lv+\rho)^{\alpha}}
\end{aligned}
 \label{eqkern1}
\end{equation}
We rewrite the  factor $\ensav{(\lv+\rho)^{\alpha}}$ on both the left and right hand sides as
\begin{equation*}
 \ensav{(\lv+\rho)^{\alpha}} = \lv\ensav{(\lv+\rho)^{\alpha-1}} + \ensav{\rho(\lv+\rho)^{\alpha-1}},
\end{equation*}
and divide the equation by $\lv\ensav{(\lv+\rho)^{\alpha-1}}$ to obtain
\begin{equation}
\begin{aligned}
  \flt \resdens(k,\lv) + \flt \resdens(k,\lv) \frac{\ensav{\rho (\lv+\rho)^{\alpha-1}}}{\lv\ensav{(\lv+\rho)^{\alpha-1}}}
   =  -\Da k^{2} \frac{1}{\lv\ensav{(\lv+\rho)^{\alpha-1}}} +
  \frac{\ft c_{0}(k)}{\lv}\left(1 + \frac{\ensav{\rho (\lv+\rho)^{\alpha-1}}}{\lv\ensav{(\lv+\rho)^{\alpha-1}}} \right).
\end{aligned}
\end{equation}
This can be written
\begin{equation}
\begin{aligned}
   \flt \resdens(k,\lv) +  \lt \kRc(\lv) \flt \resdens(k,\lv)
   =  -\Da k^{2} \lt \Kda(\lv) \flt \resdens(k,\lv) + \frac{\ft c_{0}(k)}{\lv} + \frac{\ft c_{0}(k)}{\lv} \lt \kRc(\lv).
\end{aligned}
\end{equation}
where $\lt\kRc(\lv)$ and $\lt\Kda(\lv)$ are given by
\begin{align}
\label{kernelss:supps}
\lt \kRc(\lv) = \frac{\langle \rho (\lv + \rho)^{\alpha - 1} \rangle}{\lv\langle (\lv + \rho)^{\alpha-1} \rangle}, &&
\lt\Kda(\lv) = \frac{1}{\lv\langle (\lv + \rho)^{\alpha - 1} \rangle}.
\end{align}
Inverting we obtain
\begin{align}
\resdens(x,t) - \int_0^t d t' \Kda(t - t^\prime)
\Da \frac{\partial^2 \resdens(x,t^\prime)}{\partial x^2}
 =  \int_0^t d t^\prime \kRc(t - t^\prime) \left[ c_{0}(x) - \resdens(x,t^\prime)\right] + c_{0}(x).
\label{nldrinteg}
\end{align}
Taking the time derivative, we find~\eqref{nldr1reset}
\begin{align}
\frac{\partial \resdens(x,t)}{\partial t} - \frac{\partial}{\partial t}\int_0^t d t' \Kda(t - t^\prime)
\Da \frac{\partial^2 \resdens(x,t^\prime)}{\partial x^2}
  =  -\frac{\partial}{\partial t} \int_0^t d t^\prime \kRc(t - t^\prime) \resdens(x,t^\prime) + \kRc(t) c_{0}(x).
\label{nldr1}
\end{align}
%

\subsection{Stationary densities of diffusion with disordered reset \label{supps:ctrwfpt}}

We have the density under disordered reset in the the Fourier-Laplace domain~\eqref{ctrwfptresetdens:supps}.
Multiplying by $\lv$ and taking the limit $\lv\to 0$ we obtain the stationary density
under disordered reset
\begin{equation}
  \flt \resdens(k,\lv) = \frac{\ft c_{0}(k)\ensav{\rho^{\alpha}} }
  {\ensav{\rho^{\alpha}} + k^2 \Da}.
\label{ctrwfptresetdensstat:supps}
\end{equation}
On the other hand the density under disordered decay~\eqref{genMW:supps} may be written
using~\eqref{dissurv:supps} as
\begin{equation}
  \flt p(k,\lv) = \lt p(\lv)
  \frac{\ft c_{0}(k)\ensav{(\lv+\rho)^{\alpha}} }
  {\ensav{(\lv+\rho)^{\alpha}} + k^2 \Da}.
  \label{dissurvrewrite:supps}
\end{equation}
That the hypothesis of Theorem~\ref{thm:stationaryone:supps} is satisfied can be understood as follows.
For $\psi_{\rho}(r) \sim r^{\nu-1}$ as $r\to 0$,
the expansion of $\ensav{(\lv+\rho)^{\alpha}}$ in $\lv$
has $\max\{k: k < \alpha + \nu\}$ regular terms before
the first non-analytic term. The factor $\ensav{(\lv+\rho)^{\alpha-1}}$,
which appears only in the numerator of $\lt p(\lv)$, has $k-1$ such
terms, so $\lt p(\lv)$ dominates as $\lv\to 0$ and we have
\begin{equation}
  \ft p(k,t) \sim p(t)
  \frac{\ft c_{0}(k)\ensav{\rho^{\alpha}} }
  {\ensav{\rho^{\alpha}} + k^2 \Da} \text { as } t \to \infty.
\end{equation}
The normalized density under decay $\ft c(k,t) = \ft p(k,t)/p(t)$,
is thus identical to that under reset~\eqref{ctrwfptresetdensstat:supps} as $t\to\infty$.
Eq.~\eqref{dissurvrewrite:supps}, may be Fourier-inverted
(with $c_0(x)=\delta(x)$) to obtain
\begin{align}
\label{propagator}
\lt p(x,\lv) = \frac{\lt p(\lv)}{2} {\mathcal R}(\lv) e^{- |x| {\mathcal R}(\lv)},
\end{align}
where ${\mathcal R}(\lv) \equiv \Da^{-1/2} \ensav{(\lv+\rho)^{\alpha}}^{1/2}$.

\end{document}